%% file: main.tex
\documentclass[sigconf]{acmart}

\usepackage{booktabs} 
\usepackage{listings}
\usepackage{amsmath}
\usepackage{amssymb}
\usepackage{graphicx}
\usepackage{setspace}
\usepackage{fullpage}
\usepackage{xcolor}
\usepackage{algorithm}
\usepackage{indentfirst}
\usepackage{subfigure}
\usepackage{enumitem}
\usepackage[font=normal,skip=0pt]{caption}

\setcopyright{none}

\usepackage{color}
\usepackage[normalem]{ulem}
\definecolor{mred}{rgb}{.80,.12,.30}
\definecolor{grey}{rgb}{.5,.5,.5}

\newif\ifnotes
\notestrue

\definecolor{lightgray}{rgb}{.9,.9,.9}
\definecolor{darkgray}{rgb}{.4,.4,.4}
\definecolor{purple}{rgb}{0.65, 0.12, 0.82}

\lstdefinelanguage{JavaScript}{
  keywords={typeof, new, true, false, catch, function, return, null, catch, switch, var, if, in, while, do, else, case, break},
  keywordstyle=\color{blue}\bfseries,
  ndkeywords={class, export, boolean, throw, implements, import, this},
  ndkeywordstyle=\color{darkgray}\bfseries,
  identifierstyle=\color{black},
  sensitive=false,
  comment=[l]{//},
  morecomment=[s]{/*}{*/},
  commentstyle=\color{purple}\ttfamily,
  stringstyle=\color{red}\ttfamily,
  morestring=[b]',
  morestring=[b]"
}
\acmConference[CIDR'19]{}{}{Asilomar, CA USA}

\begin{document}
\title{Kyrix: Interactive Visual Data Exploration at Scale}

\author{Wenbo Tao}
\affiliation{%
  \institution{MIT CSAIL}
}
\email{wenbo@mit.edu}

\author{Xiaoyu Liu}
\affiliation{
  \institution{Purdue University}
}
\email{liu1962@purdue.edu}

\author{{\c{C}}a{\u{g}}atay Demiralp}
\affiliation{
  \institution{MIT CSAIL}
}
\email{cagatay@csail.mit.edu}

\author{Remco Chang}
\affiliation{
  \institution{Tufts University}
}
\email{remco@cs.tufts.edu}

\author{Michael Stonebraker}
\affiliation{
  \institution{MIT CSAIL}
}
\email{stonebraker@csail.mit.edu}

\renewcommand{\shortauthors}{B. Trovato et al.}

\DeclareRobustCommand{\subhead}[1]{\noindent\textbf{#1}}

\input{abstract}


\maketitle

\input{intro}

\input{example}

\input{scalability}

\input{discussion}

\input{related}

\input{conclusion}

\bibliographystyle{ACM-Reference-Format}
\bibliography{kyrix}

\end{document}

%% file: abstract.tex
\begin{abstract}
Scalable interactive visual data exploration is crucial in many domains due to increasingly large datasets generated at rapid rates. Details-on-demand provides a useful interaction paradigm for exploring large datasets, where users start at an overview, find regions of interest, zoom in to see detailed views, zoom out and then repeat. This paradigm is the primary user interaction mode of widely-used systems such as Google Maps, Aperture Tiles and ForeCache. These earlier systems, however, are highly customized with hardcoded visual representations and optimizations. A more general framework is needed to facilitate the development of  visual data exploration systems at scale. In this paper, we present Kyrix, an end-to-end system for developing scalable details-on-demand data exploration applications. Kyrix provides developers with a declarative model for easy specification of general visualizations. Behind the scenes, Kyrix utilizes a suite of performance optimization techniques to achieve a response time within 500ms for various user interactions. We also report results from a performance study which shows that a novel dynamic fetching scheme adopted by Kyrix outperforms tile-based fetching used in earlier systems.
\end{abstract}

%% file: intro.tex
\section{Introduction}

Scaling interactive visual data exploration to massive datasets is becoming
increasingly important with the rapid generation of data across domains, from
healthcare to sciences.  It is not unusual for analysts in application domains
to deal with datasets of sizes in the order of terabytes or petabytes. Since fluid interactions help allocate human attention efficiently over data\cite{liu2014effects}, interactivity should not be compromised when exploring big datasets, which can easily overwhelm analysts. 

Details-on-demand~\cite{Shneiderman:1996:ETD}
is a common interaction pattern that arises from exploratory data analysis
practices and can be particularly effective in exploring complex datasets,
reducing the user's information load. In this interaction pattern users start
with an overview of a dataset and then zoom into a smaller subset of interest
within the dataset to examine this data patch~\cite{Pirolli:1999:Foraging},
while querying details on items within the focused region as needed.  Users
repeat the same process after zooming further into or zooming out of the
current region. However, most visual exploration systems cannot handle
very big datasets, let alone enable details-on-demand interactions. Large
data scales make it challenging to bound the interaction response times within
500ms, which is required for sustaining an interactive user
experience~\cite{liu2014effects}.

\begin{figure}[!t] 
  \centering
  \includegraphics[height=4.2cm]{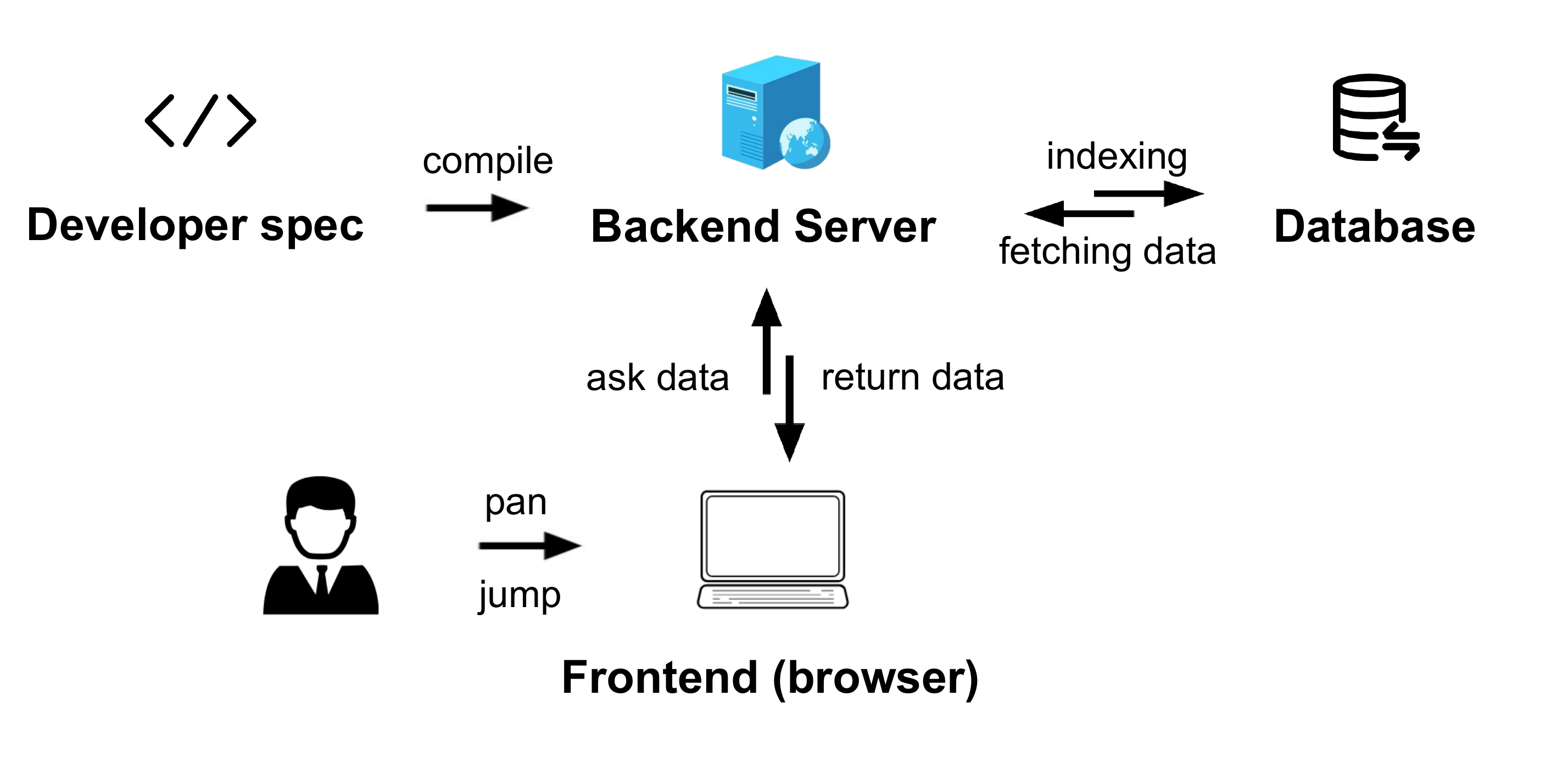}
  \caption{Architecture of Kyrix, an end-to-end system for developing scalable
  details-on-demand visualizations.\label{fig:architecture}} 
\end{figure}

\begin{figure*}[t]
  \centering
\includegraphics[height=3.6cm]{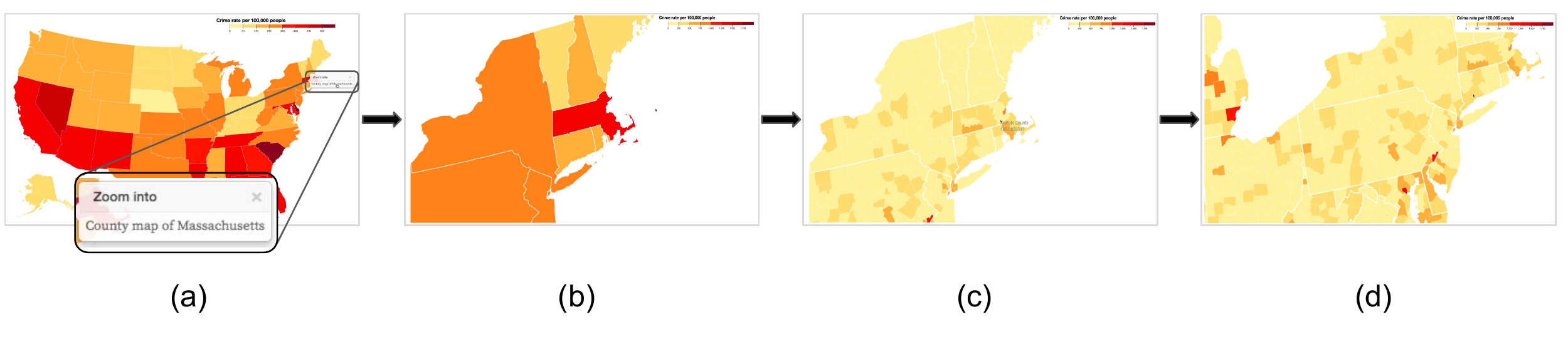}
\caption{Interactive map of crime rates in the US: (a) a state-level crime rate
  map where a user can click on a state and zoom into a county-level crime rate
  map centered at the selected state; (b) after the user chooses to zoom, Kyrix
  frontend starts a smooth zoom transition; (c) county-level crime rate map
centered at Massachusetts; (d) the user pans on the county-level map.}
  \label{fig:usmap}
\end{figure*}

Several earlier details-on-demand systems address interactivity challenges at
scale with highly-customized implementations.  Google Maps and Aperture
Tiles~\cite{aperturetiles} precompute image tiles of the entire world map at
multiple detail (zoom) levels for scalable panning and zooming.  Similarly,
imMens~\cite{liu2013immens} supports interactive panning, zooming and brushing
\& linking in binned plots by precomputing tiled data cubes.
ATLAS~\cite{Chan:2008:ATLAS} uses predictive prefetching and level-of-detail
management to improve the performance of panning and zooming interactions on
large time-series datasets.  ForeCache~\cite{Battle:2016:Forecache} also uses
predictive prefetching along with data tiling to help sustain an interactive
details-on-demand exploration of satellite images.  Although these earlier
systems use similar approaches to facilitate scalability, they are highly
customized one-off tools developed from scratch for specific  datasets. The
optimization techniques used in these systems are often inaccessible to
visualization developers at large, who are not necessarily experts in
performance optimization. Furthermore, current visualization specification
tools provide no or limited support for developers to create interactive
visualization applications at scale. 

To accelerate and improve the development of scalable visual data exploration
systems, we need general purpose tools that can help developers handle large
datasets by using effective optimization techniques (e.g. indexing, caching and
prefetching).  This warrants an integrative,
end-to-end approach to visualization specification, where performance
optimizations and data are pushed to the server side computation and DBMSs,
which can scale reasonably well with increasing data sizes.

 In this paper, we present the design of Kyrix, a novel system for developers
 to build large-scale details-on-demand visualizations. Our goal is to achieve
 both generality and scalability. Figure~\ref{fig:architecture} shows the
 architecture of Kyrix. On the developer side, we offer a concise yet
 expressive declarative language for specifying visualizations. Declarative
 designs hide execution details (e.g.  backend optimization, frontend
 rendering) from developers, so that they can focus on visual
 specification~\cite{2014-reactive-vega}. On the execution side, there are
 three main components: the compiler, the backend server, and the frontend
 renderer. The compiler parses developers' specification and performs basic
 constraint checkings. Based on the developer specification, the backend server
 then builds indexes and performs necessary precomputation. The frontend
 renderer is responsible for listening to users' activities, communicating with
 the backend server to fetch data and rendering the visualizations.

In the following, we first discuss a simple map visualization created using
Kyrix, briefly demonstrating the use of its services and declarative language.
We then introduce optimizations used by Kyrix that facilitate the development
of fluid details-on-demand interactions. Next we discuss useful
extensions to the Kyrix system along with avenues of future research. We then
put Kyrix in the context of earlier scalable visualization systems and grammars
for declarative visualization specification. We conclude by summarizing our
contributions and reiterating our vision on accelerating the development of
interactive visualizations for massive datasets.

%% file: example.tex
\section{Developing Interactive Visualizations with Kyrix} 

The goal of Kyrix is to provide an end-to-end solution for visualization
developers to create details-on-demand visualizations. To this end, Kyrix
provides a declarative language for visualization specification. 

\subsection{Kyrix Declarative Language}

 Kyrix's declarative model has two basic abstractions: \textit{canvas} and \textit{jump}. A canvas is an arbitrary size worksheet with one or more
 overlaid \textit{layers}, forming a single view showing a static visualization. A jump is a customized transition from one canvas to another.
 This model allows easy specifications of common details-on-demand interactions
 such as panning, geometric and semantic zooming \footnote{Geometric zooming refers to scaling the visualization to show different levels of details. Data type and visual encoding are unchanged. Semantic zooming, in contrast, connects different views showing related data using smooth zoom-like transitions. Data type and visual encoding can both be changed.}.
 
  
The Kyrix declarative language is data type agnostic and supports a myriad of
specific visualizations.  To render a layer, developers specify the following: 

\begin{itemize}
  \item[(1)] The data needed for the layer.  This is specified using a SQL
    query to a DBMS along with a transform function postprocessing the query
    result. Developers can use existing visualization libraries (e.g., D3 and
    Vega) to specify a desired transform function (e.g., layout transforms,
    scaling, etc). However, this transform function is not required. Developers
    still can and should transform their data outside Kyrix if it is more
    convenient. 
  \item [(2)] The location of each returned data object on the canvas. This is
    specified using a placement function.
  \item [(3)] A rendering function that converts a canvas object to pixels 
    on the screen.  Kyrix's rendering functions can be written using 
    lower-level visualization specification libraries such as D3.
\end{itemize}

A jump transition can be established simply by specifying a \textit{from}
canvas, a \textit{to} canvas and a transition type (right now it can be geometric
zoom, semantic zoom or both). It can also be customized in many ways. For example,
developers can specify a subset of objects on the \textit{from} canvas that can
trigger this jump. For more details on the language, interested readers can
refer to our developer
manual\footnote{https://github.com/tracyhenry/Kyrix/blob/master/compiler/README.md}. 

\subsection{Example: Map of US Crime Rate} 
We now describe an interactive application created using Kyrix.  The example
visualizes the US crime rates per state and county (Figure \ref{fig:usmap}).
There are two canvases in this application. The initial canvas in
Figure~\ref{fig:usmap}a shows a map of the state-level crime rate.  Users can
click on a state and zoom into a second, pannable canvas that shows the crime
rates at the county level (Figure \ref{fig:usmap}c).  In the current
implementation, developers are expected to write specifications in Javascript.
Figure~\ref{fig:usmapsnippet} shows a snippet for the example application.  An
application object (Line 2) is constructed by specifying the application name
and a configuration file containing information such as the underlying DBMS.

The state map canvas is specified in Lines 5\textasciitilde21. This canvas
contains two overlaid layers: a static legend layer (lines 13\textasciitilde15)
and a pannable state border layer (lines 18\textasciitilde21). Each layer is
specified using an identifier of a data transform (lines 9 and 10) and a
boolean value indicating whether this layer is static (Lines 13 and 18). Static
layers do not need to be re-rendered when user pans. So in this case when user
pans, the legend will stay unchanged in the upper right-hand corner, overlaid
on the state border layer. The county map canvas is also similarly specified.
In Figure~\ref{fig:usmapsnippet} we leave out the specification of the county
map canvas along with the transform, rendering and placement functions due to
limited space. 

A jump transition from the state canvas to the county canvas is defined in line
36. In the constructed jump object, the first two arguments respectively
identify the state and county canvases. The third argument specifies the jump
type.  The rest of the arguments are used to customize the jump transition.
To complete the specification of the application, developer would also specify
an initial canvas and a viewport center (line 39). 

\begin{figure}[t] 
\centering 
\begin{lstlisting}[frame = none] 
// construct an application object 
var app = new App("usmap", "config.txt");

// ================== state map canvas =================== 
var stateMapCanvas = new Canvas("statemap"); 
app.addCanvas(stateMapCanvas);

// add data transforms 
stateMapCanvas.addTransform(transforms.emptyTransform);
stateMapCanvas.addTransform(transforms.stateMapTransform);

// static legend layer 
var stateMapLegendLayer = new Layer("empty", true);
stateMapCanvas.addLayer(stateMapLegendLayer);
stateMapLegendLayer.addRenderingFunc(renderers.stateMapLegendRendering);

// state border layer 
var stateBorderLayer = new Layer("stateMapTrans", false);
stateMapCanvas.addLayer(stateBorderLayer);
stateBorderLayer.addPlacement(placements.stateMapPlacement);
stateBorderLayer.addRenderingFunc(renderers.stateMapRendering);

// ================== county map canvas =================== 
...

// =================== state -> county ==================== 
var selector = function (row, layerId) {
    return (layerId == 1);
};
var newViewport = function (row) {
    return [0, row[1] * 5 - 1000, row[2] * 5 - 500];
};
var jumpName = function (row) {
    return "County map of " + row[3];
};
app.addJump(new Jump("statemap", "countymap", "geometric_semantic_zoom", selector, newViewport, jumpName));

// set initial canvas 
app.initialCanvas("statemap", 0, 0);

\end{lstlisting} 
\caption{A Javascript snippet of the US crime rate map example.\label{fig:usmapsnippet}}
\end{figure}

%% file: scalability.tex
\section{Interactivity in Kyrix}

In general, the interactivity problem in Kyrix is to achieve a 500 ms response time to the following user interactions: (1) A pan to a different location on the same canvas and (2) a jump to a different canvas. 


In Section \ref{subsec:datafetching} we discuss how Kyrix fetches data in
response to user interactions. Then in Section \ref{subsec:perfguideline}, we
give some general guidelines that will assist with achieving our goal. Lastly,
Section \ref{subsec:perfnumber} gives some end-to-end performance numbers
concerning achieving our goal. We will discuss in Section \ref{sec:discussion}
other performance options. If accepted, we would expect to present this work
primarily as a demo.

\subsection{Data Fetching} \label{subsec:datafetching}
As user performs one of the operations (pan or jump), or when an application is
first loaded, Kyrix's frontend communicates with the backend to retrieve the
data needed to render the viewport. Like previous systems (e.g., ForeCache
\cite{Battle:2016:Forecache}), Kyrix employs both a frontend cache and a backend cache. If there is a cache miss in both, Kyrix backend will talk to the
backing DBMS to fetch data. In this data fetching process, we identify two
important factors that can affect Kyrix's performance:(1)\textit{fetching
granularity} and (2)\textit{database design and indexing}. In the following, we
describe these two factors in detail. 

\begin{figure}[!t]
 \centering
 \subfigure[Static tiles]{
\includegraphics[height=2.5cm]{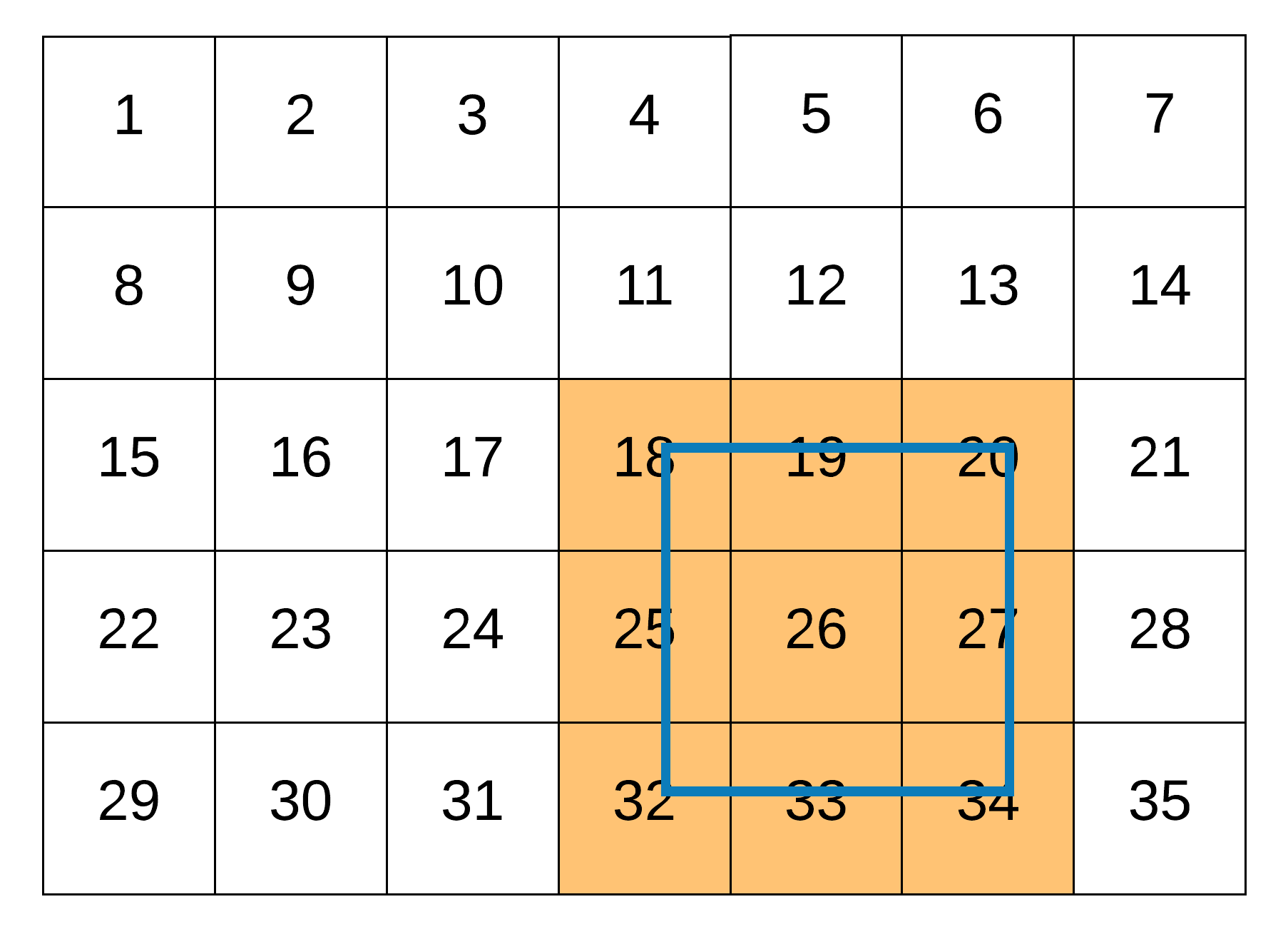}
\label{fig:tiling}}
\subfigure[Dynamic boxes]{
\includegraphics[height=2.5cm]{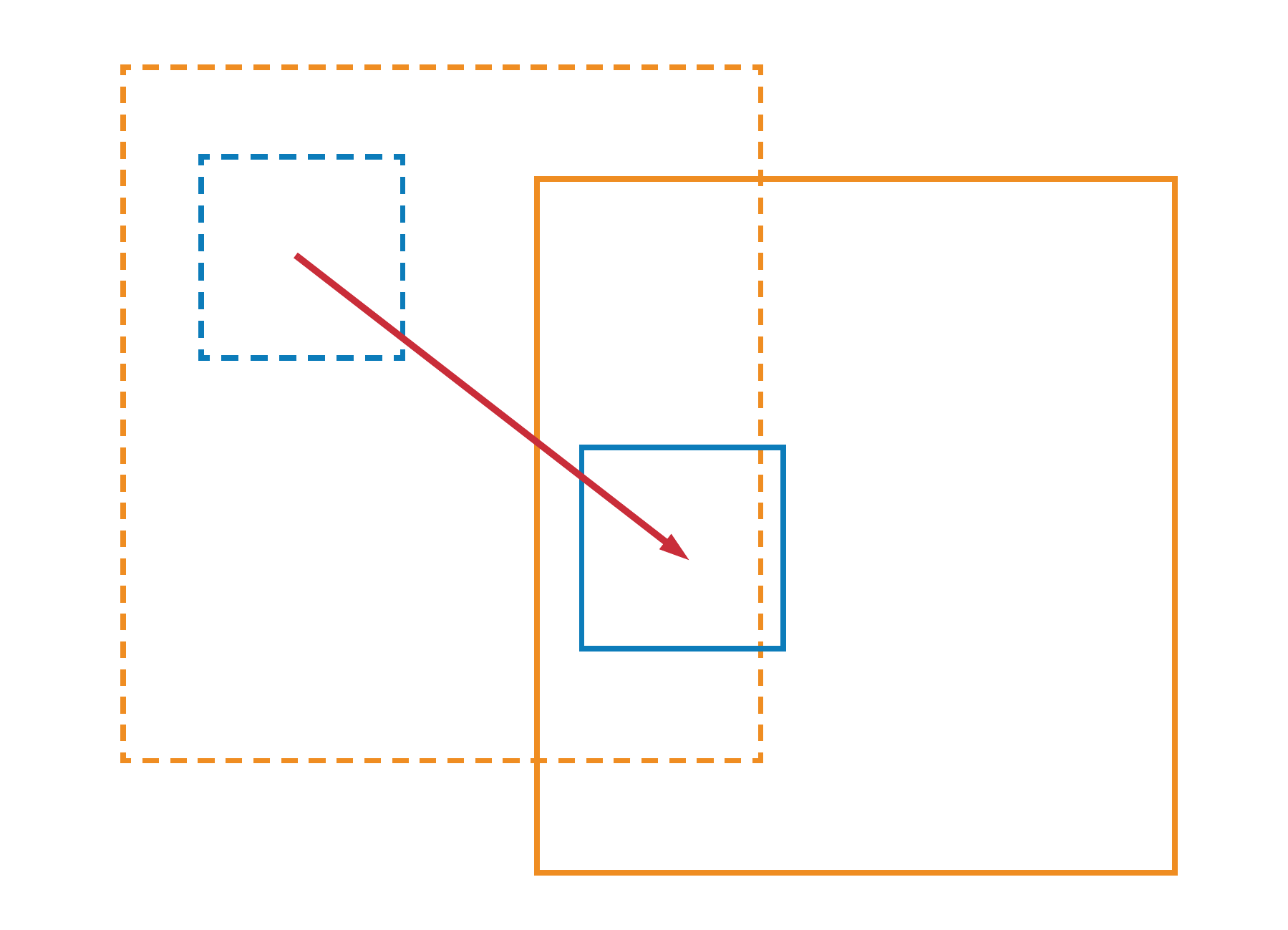}
\label{fig:dbox}}
\vspace{1em}
\caption{An illustration of two fetching granularities. (a) A canvas is partitioned into 35 tiles. Viewport is the blue rectangle. Tiles in orange are fetched. (b) Blue rectangles are viewports. Orange rectangles are what are actually fetched. Dashed-line rectangles are before the user pans. Solid-line rectangles are after the user pans. }
\end{figure}

\subhead{Fetching Granularity.}
The standard wisdom, as applied in Google Maps, ForeCache\cite{Battle:2016:Forecache} and Aperture Tiles\cite{aperturetiles},
is to decompose a canvas into fixed-size \textit{static tiles} (Figure
\ref{fig:tiling}). The frontend then requests the tiles that intersect with the
given viewport. Every tile is individually fetched and rendered. Kyrix
currently supports static tiling.  Kyrix also contributes a novel fetching
granularity, \textit{dynamic boxes}.  Dynamic box fetching amounts to
requesting a box that contains the given viewport (Figure \ref{fig:dbox}). We
call this enclosing box a dynamic box because its size and location changes
dynamically.  Whenever the viewport moves outside the current box, frontend
sends the current viewport location to backend and requests a new box. There
are numerous ways to calculate a box, e.g., a box centered at the viewport
center having width (height) 50\% larger than the viewport width (height).  We
expect dynamic boxes to outperform static tiles for the following reasons:
\begin{itemize}
  \item[(1)] compared to large tiles, dynamic boxes fetch less data;
  \item[(2)] compared to small tiles, dynamic boxes require fewer 
    frontend-backend requests in general;
  \item[(3)] in cases where data is not uniformly distributed, dynamic boxes
    can adjust their sizes and locations based on data sparsity, incurring 
    much fewer network and database trips than static tiles. 
\end{itemize}

In Section \ref{subsec:perfnumber}, we use two simple box calculation
algorithms to experimentally show that dynamic boxes are a more performant
option than static tiles. We leave an in-depth performance study as future
work. 

\subhead{Database Design and Indexing.} \label{sec:dbdesign}
We now describe two database designs along with two indexing schemes that we
use to support static tiles and dynamic boxes.  Our first database design maps
tuples to static tiles and has two tables. The first table is a record table containing all the raw data
attributes in addition to an auto-increment \textsf{tuple\_id} attribute. The
second table contains two columns \textsf{tuple\_id} and \textsf{tile\_id}.
Each record in this table corresponds to  a tuple that overlaps a tile. Kyrix
backend uses placement functions specified by developers to precompute the
second table. We then build Btree/hash indexes on the \textsf{tuple\_id} column
of the first table and the \textsf{tile\_id} column of the second table. At
runtime, tile queries are answered by joining these two tables on the
\textsf{tuple\_id} column. 

Our second database design is based on spatial index in PostgreSQL. In addition
to raw data attributes, we store a \textsf{bbox} attribute representing the
bounding box of a tuple on a canvas\footnote{We assume records are generally rendered bigger than a single pixel. This bounding box information is derived
from the placement functions specified by developers.}. We then build a spatial
index on the \textsf{bbox} column. Using this design, queries that request tuples whose bounding boxes intersect with a given rectangle should run fast.
Therefore, this design can be used by both static tiles and dynamic boxes. 


\subsection{Performance Hygiene} \label{subsec:perfguideline}

\subhead{Parallelism.} We can apply parallelism to improve the data management
in Kyrix. All data and metadata (canvas definitions, etc.) are stored in and
retrieved from the DBMS. 
Although the performance experiments in the next section use PostgreSQL, it
would be prudent to replace the DBMS with a parallel one if performance
requirements warrant a switch.  Currently, rendering is performed by a separate
process on a separate CPU in the frontend. This operation can also be easily
parallelized. Lastly, each concurrent Kyrix application is run in a separate
process, since there is no interaction between them, except through the DBMS.
Right now, Kyrix applications function like a read-only browsers. Future
releases will extend Kyrix to allow editing updates, which can be supported by
DBMS concurrency control.

\subhead{Application Design.} Managing visual density on the screen, which 
can overwhelm users as well as the client (e.g., the browser) resources, 
is an important concern in visualization of large datasets.  
Application design must deal with what canvases exist and how to 
put data onto these canvases so that visual density is not too high.  

\subhead{Separability.} Recall in Section \ref{sec:dbdesign}, we describe how
Kyrix precomputes database tables and indexes to ensure data fetching speed.
However, when data is huge or the SQL query corresponding to a canvas layer is
complex, this precomputation process can take a long time.  We identify a
common case where this precomputation process can be avoided: the ($x$, $y$)
placement of objects are directly raw data attributes, or some simple scaling
of raw data attributes. In these \textit{separable} cases, if we assume DBAs
have built spatial indexes on relevant raw data attributes when data is first
loaded into the DBMS, we do not have to precompute the tables described in
Section \ref{sec:dbdesign}. For separable cases, we provide developers with the
option to specify the relevant attributes so that precomputation can be skipped
by Kyrix.  There are cases where this requirement cannot be met, i.e., the
placement of an object depends on multiple data attributes or the placements of
other objects. We call these cases non-separable. Pie chart is an example. 



\begin{figure}[!t]
    \centering
    \includegraphics[width=7.7cm]{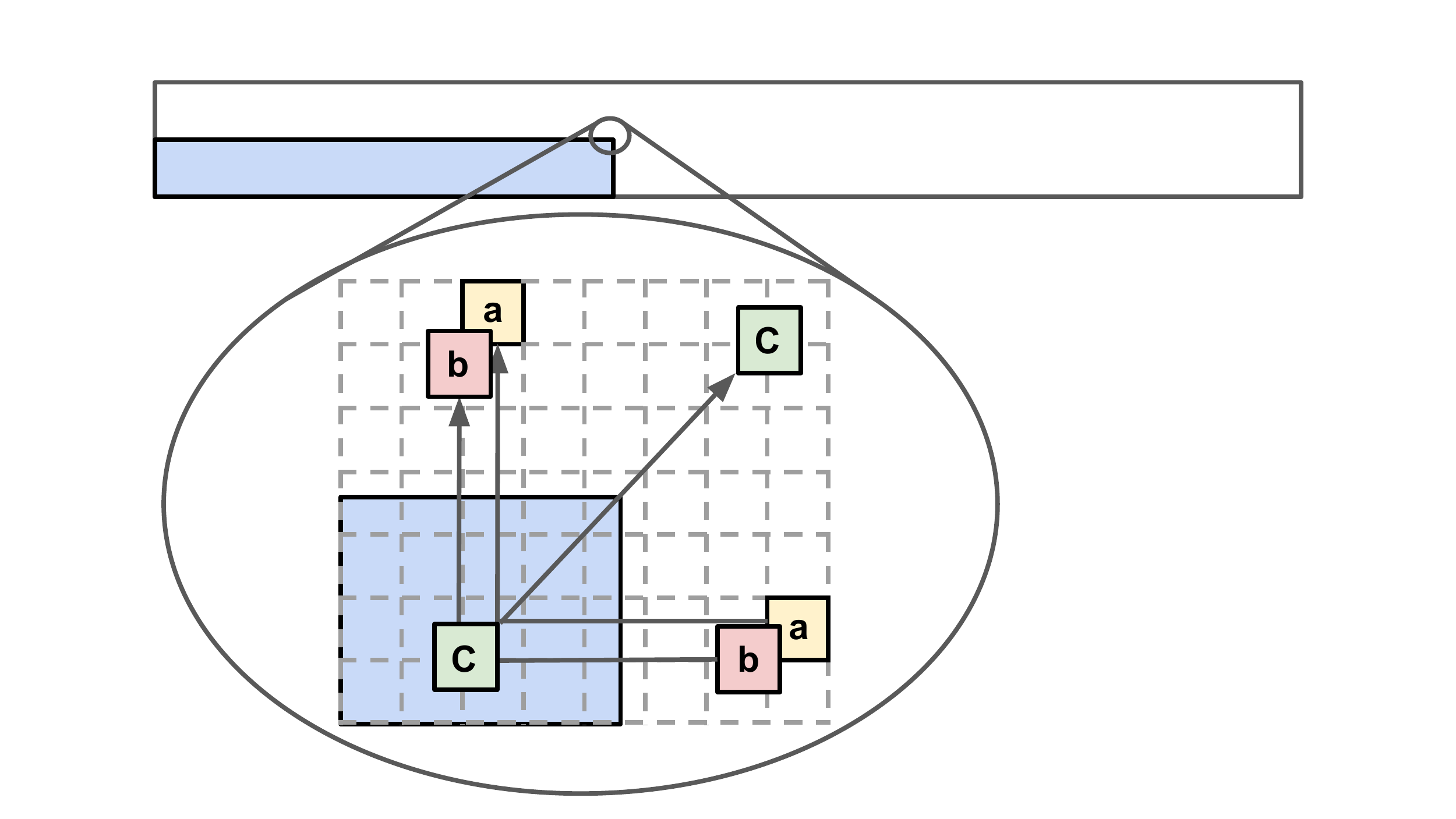}
    \caption{Viewport movement traces used in our experiments. Blue shaded area is the dense area in the dataset \textsc{Skewed}. Dotted lines are the boundaries of tiles with size 1,024.\label{fig:trace}}
    \vspace{-0.75em}
\end{figure}

\subsection{Initial Performance Experiments} \label{subsec:perfnumber}
We conducted performance experiments on two synthetic datasets using three
viewport movement traces. The goal of these experiments is to study the
characteristics of the two fetching granularities when combined with different
database designs. All experiments are done on an AWS EC2  m4.2xlarge instance
with 8 cores and 32GB RAM. PostgreSQL 9.3 is the backing DBMS. 

\subhead{Datasets.} We used two synthetic datasets, \textsc{Uniform} and
\textsc{Skewed}. In \textsc{Uniform}, there are 100M random dots evenly
distributed on a 1M$\times$0.1M canvas. In \textsc{Skewed}, 80M dots lie in
20\% of the canvas area (a 0.4M$\times$0.05M rectangle) and 20M dots lie in the
rest of the canvas. \textsc{Skewed} corresponds to the likely scenario when
objects are distributed unevenly on a canvas.

\subhead{Viewport Movement Traces.} In our experiments we use three viewport movement traces illustrated in Figure~\ref{fig:trace}.  

\begin{itemize}
    \item[(a)] The viewport is always aligned with tile boundaries. It horizontally moves leftwards six steps (the length of a tile) then vertically up six steps.
    \item[(b)] The viewport is never aligned with tiles. It also horizontally moves leftwards six steps (the length of a tile) then vertically upwards six steps. 
    \item[(c)] The viewport moves diagonally from bottom left to top  right. There are six steps in total. 
\end{itemize}
\subhead{Fetching schemes.} 
We evaluated the following fetching schemes.
\begin{itemize}[leftmargin=0pt]
  \item[]\textit{Dbox}: Dynamic boxes with spatial index. The box fetched
    is exactly the viewport in each step.
  \item[]\textit{Dbox 50\%}: Dynamic boxes with spatial index. The box
    fetched is 50\% larger than the viewport.
  \item[]\textit{Tile spatial}: Static tiles with spatial index (three
    tile sizes tested: 256, 1,024 and 4,096).
   \item[]\textit{Tile tuple-tile mapping}: Static tiles with tuple-tile
    mapping (tile size 1,024 tested). Btree index is used on \textsf{tuple\_ID}
    and \textsf{tile\_ID} columns.
\end{itemize}

\begin{figure}
\centering
\includegraphics[width=7.5cm]{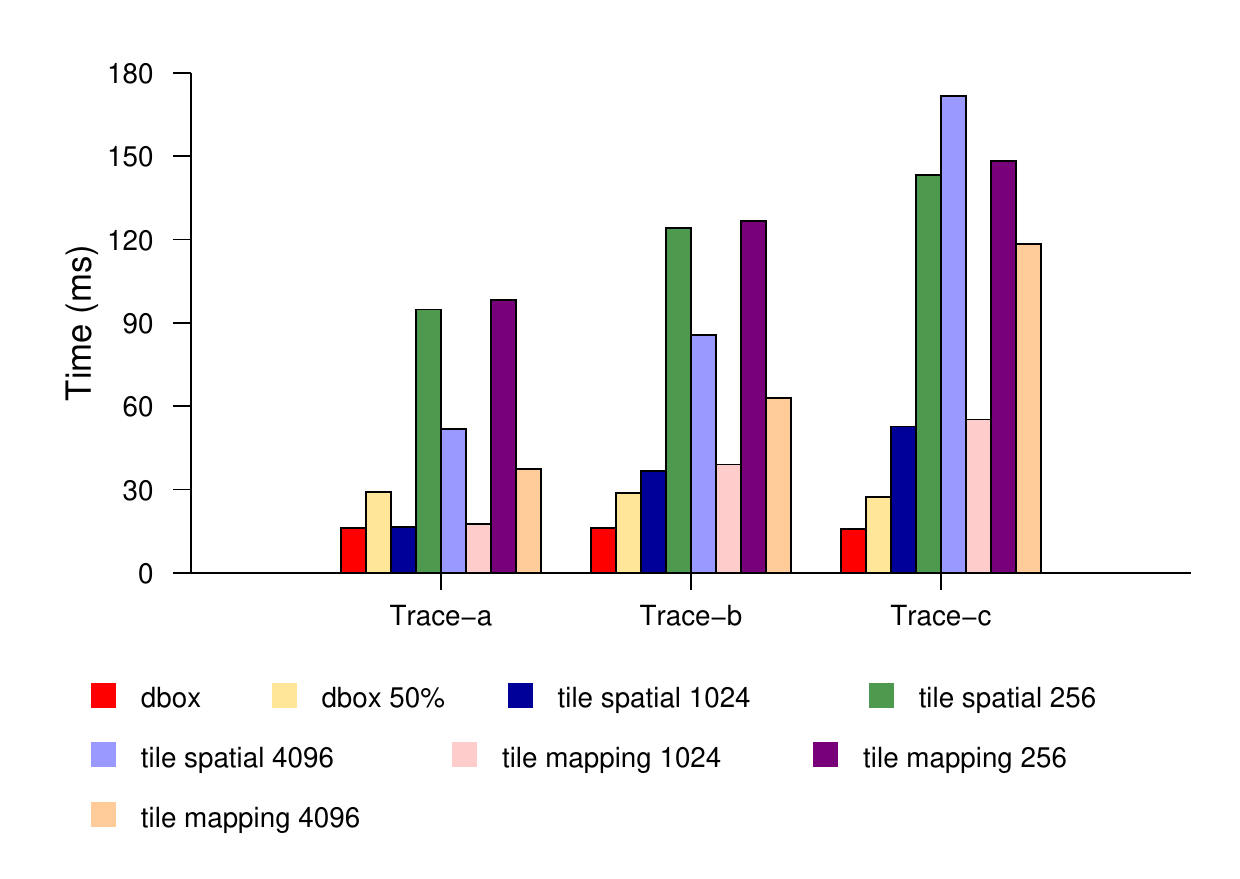}
\caption{The average response times of dynamic box and static tiling on uniformly distributed data.}
\label{fig:uniform}
\end{figure}
\begin{figure}
\centering
 \includegraphics[width=7.5cm]{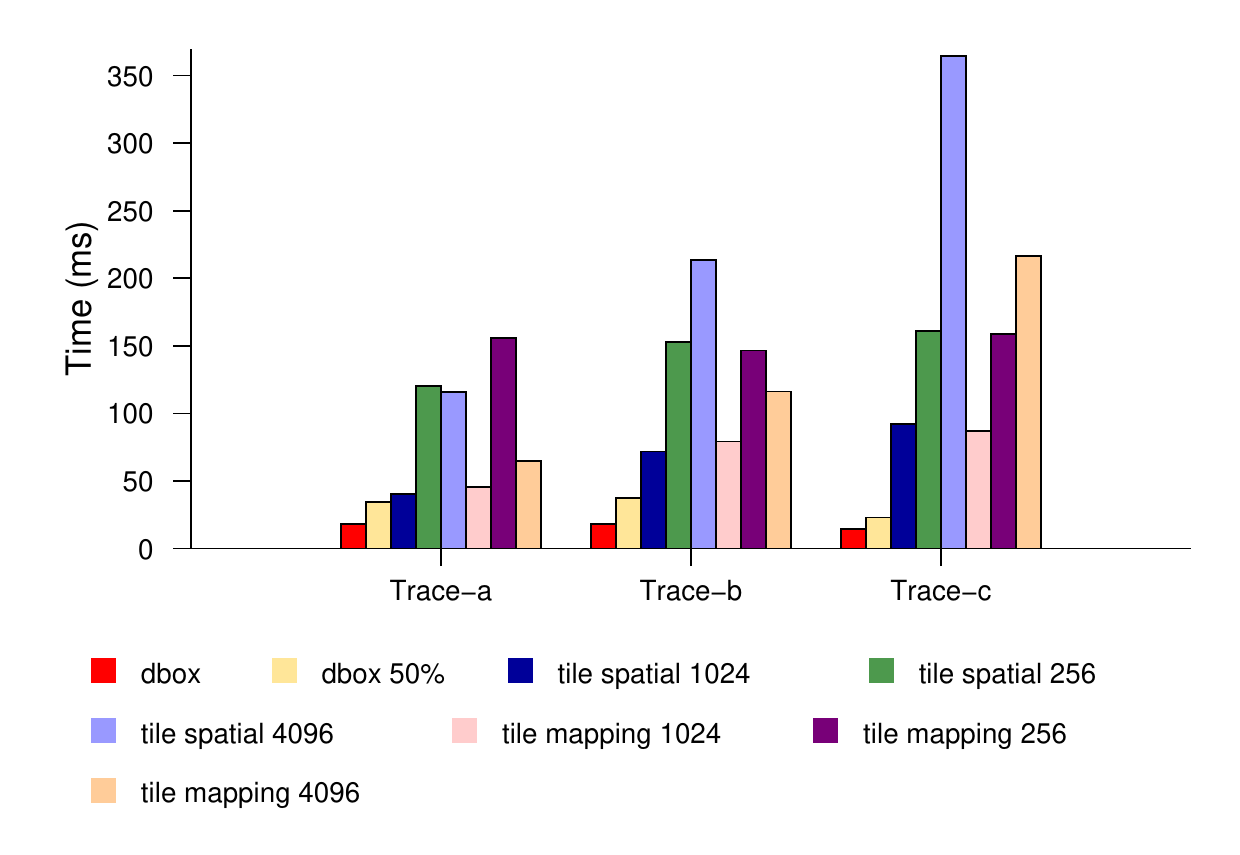}
\caption{The average response times of dynamic box and static tiling on skewed data. }
\label{fig:skew}
\end{figure}

\subhead{Results.} We measured the average response time (per step) of all fetching schemes on three traces. Average results over three runs are shown in Figures \ref{fig:uniform} and \ref{fig:skew}. We have the following observations:

\begin{itemize}
  \item[(1)]\textit{Dbox} has the best overall performance on both \textsc{Uniform} and \textsc{Skewed}. The reasons are twofold. First, it fetches the least amount of data needed to render the viewport. Second, compared to small tiles, it issues much fewer queries.
  \item[(2)] \textit{Tile 1,024 spatial} has competitive performance on trace-a, and is even better than \textit{Dbox 50\%}. This is because the viewport completely aligns with tile boundaries in trace-a.
 \item[(3)] \textit{Tile 4,096} and \textit{256 spatial} have the worst performances. This is expected since the tile size 4,096 fetches more data than other fetching schemes and the tile size 256 issues more queries than other fetching schemes. 
\end{itemize}


%% file: discussion.tex
\section{Discussion and Future Work} \label{sec:discussion}

Previous work \cite{Battle:2016:Forecache} has studied prefetching data ahead
of the user's interaction.  Specifically, both momentum-based and
semantic-based prefetching were considered in a tiling context.  To determine
what to prefetch, semantic-based prefetching uses the similarity to recently
viewed data in data characteristics (e.g., distribution).   Whereas,
momentum-based prefetching takes the user's recent movements (e.g., pan and
zoom) into account to that end. We plan to evaluate the effectiveness of
momentum-based prefetching in the context of dynamic boxes.  Our future work
will also study caching options for Kyrix.  Caching and prefetching are
challenging given the jump operation, and will be  more challenging by the
extension of Kyrix to support coordinated views. 

Currently, we are collaborating with a neurology group at Massachusetts General
Hospital (MGH), which we anticipate motivating various future 
extensions of Kyrix. Our collaborators want to be able to interactively explore 50
terabytes of electroencephalogram (EEG) data collected from sleeping subjects.
They want three different views of the data, a temporal view, a spectral view
and a composite clustering view, to be coordinated. For instance, movement in
the temporal view should cause an appropriate change in the spectral view.
Hence, Kyrix must be extended to support multiple canvases on the screen
simultaneously and to have pan/zoom operations in one canvas cause desired
actions in other canvases. In addition, MGH wants an update model for Kyrix so
they can edit and tag relevant data. Fifty terabytes will require a parallel
multi-node DBMS to achieve our performance goals. 

Lastly, we envision Kyrix as an integrated environment for developing scalable
visualization applications. To this end, e.g, we plan to work on an
``application by example'' interface, whereby a user can drag and drop screen
objects, and Kyrix can learn to automatically generate the location function
(and perhaps other parts of the application).

%% file: related.tex
\section{Related Work}
Kyrix is related to prior efforts in scalable visualization 
systems and declarative visualization specification.  

\vspace{-.5em}
\subsection{Scalable Visualization Systems}
Earlier research has proposed methods for scalable interactive data analysis
that fall into one of the two categories in general: precomputation and
sampling~\cite{hellerstein15}. Precomputation, which traditionally referred to
processing data into formats such as prespecified tiles or cubes, has been the
prevalent approach to interactively answer queries via zooming, panning,
brushing and linking.  
Google Maps precompute image tiles for multiple zoom
layers  to support scalable panning and zooming.  Extending the tiling idea to
structured data, imMens~\cite{liu2013immens} computes multivariate data
tiles in advance along with projections corresponding to materialized database and
performs fast ``roll ups''and rendering on the GPU.
Nanocubes~\cite{lins2013nanocubes}  stores and queries multi-dimensional
aggregated data at multiple levels of resolution in memory for visualization.
Hashedcubes~\cite{Pahins:2017:Hashedcubes} improves on the memory footprint and
implementation complexity of Nanocubes with an incurred cost of longer query
times. ForeCache~\cite{Battle:2016:Forecache} uses data tiling together with predictive
prefetching and in-memory caching to enable scalable panning and zooming for
visualizations of array-based datasets. 
When precomputation is not possible (e.g., queries are not known in advance),
sampling, often combined with precomputation, and online
aggregation~\cite{hellerstein1997online,agarwal2013blinkdb,Battle:2013:Scalar}  are 
used to improve user experience.  

Kyrix precomputes database indexes and uses novel data
fetching mechanisms to efficiently respond to pan and zoom interactions. 
Kyrix's new dynamic-box fetching together with spatial index outperforms tile-based fetching used in earlier 
systems. To ensure the 500ms response time, Kyrix also adopts predictive
prefetching and caching techniques~\cite{Chan:2008:ATLAS,Battle:2016:Forecache}.  


 
\vspace{-.5em}
\subsection{Declarative Visualization Specification}

Earlier research  proposes declarative grammars over data as well as visual
encoding and design variables to specify visualizations.  In a seminal work,
Wilkinson introduces a grammar of graphics~\cite{wilkinson:book99} and its
implementation (VizML), forming the basis of the subsequent research on
visualization specification. Drawing from Wilkinson's grammar of graphics,
Polaris~\cite{Stolte_2002} (commercialized as Tableau) uses a table algebra,
which later evolved to VizQL~\cite{hanrahan2006vizql}, the underlying
representation of Tableau visualizations. Wickham introduces
ggplot2~\cite{wickham2010layered}, a widely-popular package in the R
statistical language,  based on Wilkinson's grammar.  Similarly,
Protovis~\cite{bostock2009protovis}, D3~\cite{bostock2011d3},
Vega~\cite{satyanarayan2016vega}, Brunel~\cite{wills2017brunel}, and
Vega-Lite~\cite{satyanarayan2017vegalite} all provide grammars to declaratively
specify visualizations. 

Kyrix's declarative grammar differs from these earlier efforts by 
providing constructs for specification of scalable interactive visualizations
and  integrating visualization specification with a server-side processing and scalable data management for performance optimization.

%% file: conclusion.tex
\section{Conclusion}
The current practice of purpose-built scalable visualization tools is itself
not scalable under the fast growth of large datasets across domains. To  
accelerate the development pace of interactive visualization systems 
at scale, we need to make it easier for developers to access scalable data management models as well as performance optimizations needed for sustaining
interactive rates. In this paper, we present the design of Kyrix, a novel
end-to-end system for developers to build interactive,  details-on-demand 
visualizations at scale.  Kyrix enables   developers to declaratively specify
visualizations, while utilizing  Kyrix's suite of optimizations and data management model.  
Kyrix also contributes a novel dynamic fetching scheme that 
outperforms tile-based fetching  common to existing systems. 
